# Consensus Power Inequality:
# A Comparative Study of Blockchain Networks

Kamil Tylinski[1,2], Abylay Satybaldy[2] and Paolo Tasca[1,2]

[1] UCL Centre for Blockchain Technologies, London, UK
[2] Exponential Science Foundation, London, UK

**Abstract**

The distribution of consensus power is a cornerstone of decentralisation, influencing the security, resilience, and fairness of blockchain networks while ensuring equitable impact among participants. This study provides a rigorous evaluation of consensus power inequality across five prominent blockchain networks—Bitcoin, Ethereum, Cardano, Hedera, and Algorand—using data collected from January 2022 to July 2024. Leveraging established economic metrics, including the Gini coefficient and Theil index, the research quantitatively assesses how power is distributed among blockchain network participants. A robust dataset, capturing network-specific characteristics such as mining pools, staking patterns, and consensus nodes, forms the foundation of the analysis, enabling meaningful comparisons across diverse architectures. Through an in-depth comparative study, the paper identifies key disparities in consensus power distribution. Hedera and Bitcoin demonstrate more balanced power distribution, aligning closely with the principles of decentralisation. Ethereum and Cardano demonstrate moderate levels of inequality. However, contrary to expectations, Ethereum has become more concentrated following its transition to Proof-of-Stake. Meanwhile, Algorand shows a pronounced centralisation of power. Moreover, the findings highlight the structural and operational drivers of inequality, including economic barriers, governance models, and network effects, offering actionable insights for more equitable network design. This study establishes a methodological framework for evaluating blockchain consensus power inequality, emphasising the importance of targeted strategies to ensure fairer power distribution and enhancing the sustainability of decentralised systems. Future research will build on these findings by integrating additional metrics and examining the influence of emerging consensus mechanisms.

## 1. Introduction

Distributed ledger technology (DLT) has fundamentally reshaped how we think about trust, transparency, and decentralisation in digital systems. By enabling a decentralised and tamper-resistant record of transactions, blockchain technology[1] has powered the rise of cryptocurrencies, decentralised finance (DeFi), and various other applications. At its core, decentralisation is regarded as a key virtue of blockchain systems, promising resilience against single points of failure, improved security, and a fairer distribution of power among network participants [1]. However, achieving and maintaining decentralisation is a complex challenge, as blockchain networks must balance performance, scalability, and security. Among the many aspects of decentralisation, the distribution of consensus power stands out as a critical metric for evaluating the robustness and equity of these systems.

Consensus power refers to the influence or control participants exert in reaching agreements and validating transactions within a blockchain network. This encompasses both crypto holders, who often participate in governance and staking in Proof-of-Stake (PoS) systems, and network node operators or validators, who play an active role in transaction validation and block production in Proof-of-Work (PoW) systems. The distribution of this power can vary significantly across blockchain networks depending on their consensus mechanisms, such as PoW and PoS, which prioritise different operational and economic incentives [2]. While PoW networks rely on computational resources for consensus, PoS systems depend on the staking of cryptocurrency, leading to different patterns of power concentration.

Inequalities in consensus power distribution have significant implications for blockchain networks. Excessive concentration of power among a few participants undermines decentralisation, creating centralisation risks that could compromise network security, reduce trust, and introduce systemic vulnerabilities [3]. This centralisation contradicts the foundational principles of blockchain, which emphasise distributed authority and community-driven governance. As blockchain technology evolves and networks compete to achieve wider adoption, ensuring an equitable distribution of consensus power remains an essential challenge.

This paper examines consensus power inequality as a key metric for understanding decentralisation within blockchain networks. By adopting established economic inequality metrics, we quantitatively evaluate how consensus power is distributed among participants. This study focuses on five prominent blockchain networks—Bitcoin, Ethereum, Cardano, Hedera and Algorand—spanning different consensus mechanisms and operational designs. By analysing data from January 2022 to July 2024, we aim to identify trends and disparities in consensus power distribution, providing actionable insights for network designers and policymakers.

---
[1] We note that the terms 'DLT' and 'blockchain' will be used interchangeably throughout this paper.

The primary contributions of this study are as follows:

- **Quantitative assessment:** The study applies established economic inequality metrics to evaluate consensus power distribution across blockchain networks, providing a rigorous analysis of the degree of inequality and its implications for decentralisation.
- **Comprehensive dataset:** Utilising a robust dataset capturing consensus power allocation across diverse blockchain networks, this study integrates network-specific characteristics to enable a more nuanced examination of inequality trends.
- **In-depth comparative analysis:** By conducting a detailed comparison of PoW and PoS blockchains, the study identifies key disparities in consensus power inequality, uncovering structural and operational factors that contribute to unequal power distribution and highlighting potential opportunities for more equitable network design.

This work builds upon prior studies that have analysed decentralisation using layered frameworks and quantitative metrics, while addressing a gap in the literature by focusing on consensus power distribution across multiple blockchain networks with varying consensus mechanisms. By comparing these networks, this study not only provides insights into the current state of decentralisation but also offers guidance for designing future blockchain systems that align more closely with the principles of equity and resilience.

## 2. Related Work

Decentralisation is widely acknowledged as a core virtue of blockchain systems. Several studies have proposed frameworks to compare decentralisation across blockchain networks, providing valuable insights into how different systems distribute power, influence, and participation. For instance, authors in [3] introduced a layered framework to assess decentralisation, highlighting consensus as a critical layer alongside others, such as network and governance. Similarly, the work in [4] categorised centralisation factors across various dimensions, including the consensus mechanism, emphasising its role in determining the distribution of computational power and decision-making authority. These studies underscore the importance of examining consensus as a foundational component when evaluating blockchain decentralisation and offer a structured basis for comparative analysis.

Recent studies have increasingly adopted economic inequality metrics to evaluate wealth decentralisation across blockchain networks [5, 6, 7]. The analysis presented in [5] conducted a comprehensive econometric evaluation of wealth distribution in eight major cryptocurrencies, including Bitcoin and Ethereum. By applying the Gini coefficient and Nakamoto index, the study highlighted significant parallels between wealth inequality in crypto economies and traditional financial systems, emphasising persistent centralisation in these networks. In addition, the research in [6] provided a detailed overview of wealth decentralisation across multiple blockchain networks, including both Layer 1 and Layer 2 systems. This study quantified wealth distribution and introduced a novel group-based HHI metric, revealing significant disparities across blockchain networks. Building on these approaches, this research evaluates consensus power distribution across blockchain networks using economic inequality metrics.

The exploration of consensus power distribution within blockchain networks has become a key focus for understanding and improving decentralisation. However, much of the recent literature on consensus power decentralisation has concentrated on Bitcoin [8, 9, 10], highlighting trends toward centralisation due to a small number of mining pools dominating the network by controlling the majority of its computational resources. From a different perspective, another study [11] revealed that nodes linked to major Bitcoin mining pools tend to have higher mining capacities than others, even though they represent a small fraction of the network. Few studies have measured and compared the level of consensus power decentralisation across multiple blockchains using quantitative metrics. For example, the study in [12] examined decentralisation in Bitcoin and Ethereum through various metrics and levels of granularity, focusing on aspects such as mining power distribution and node diversity. However, this work was restricted to two blockchains with similar consensus mechanisms, as Ethereum still utilised PoW at the time. In contrast, our study expands the scope by analysing consensus power inequality across five blockchains with diverse consensus mechanisms, providing a more comprehensive and comparative perspective.

## 3. Consensus Power Inequality in Blockchain Networks

Consensus power inequality arises when a small number of participants in a blockchain network exert a disproportionately large influence over the consensus process, leading to centralisation. This imbalance undermines the core principles of decentralisation and reduces the network's overall security, fairness, and resilience. Understanding the causes and implications of consensus power inequality is crucial for designing blockchain systems that uphold the foundational values of decentralisation.

One significant form of consensus power inequality is the concentration of mining or staking power. In PoW systems such as Bitcoin, mining power often becomes concentrated among a small group of large mining entities or pools. These entities control substantial amounts of hash power, giving them a dominant role in the consensus process. Similarly, in PoS systems, validators are selected based on their staked tokens, resulting in a concentration of consensus power among participants with large holdings. This creates a scenario where a few stakeholders wield most of the decision-making power, mirroring the centralisation challenges faced in PoW systems [13].

Governance mechanisms can further exacerbate consensus power inequality. Many blockchain networks use governance models that assign decision-making authority based on token ownership. While this approach aligns incentives, it can also result in wealthier participants or large token holders dominating governance decisions. This concentration of power in the hands of a few individuals or entities can stifle the broader community's voice and potentially harm the network's decentralisation [14].

Economic barriers also play a significant role in influencing consensus power distribution and as such should be considered when considering the use case of the blockchain in a real-world environment. In PoW systems, the high cost of mining hardware and electricity often restricts participation to those with substantial financial resources. Although PoS systems lower these barriers by relying on token ownership for staking, they can still favor wealthier participants who hold large quantities of the native cryptocurrency. This financial disparity perpetuates risks of centralisation [4].

Network effects further contribute to the concentration of consensus power. Early adopters of blockchain technology often gain significant resource advantages, allowing them to preserve their dominance over time. Additionally, large operations benefit from economies of scale, reducing their costs and reinforcing their influence within the network [15].

Technological disparities also impact consensus power distribution. Participants with access to superior hardware, optimised software, and enhanced network connectivity can validate transactions more efficiently, gaining a competitive edge. These technological advantages can create disparities, enabling certain participants to consolidate their influence over the consensus process [16].

Market dynamics, such as consolidation and speculative investment, exacerbate consensus power inequality. Competitive pressures can lead smaller entities to merge with or be acquired by larger ones, further concentrating power. Additionally, large investors can acquire significant token holdings, exerting substantial influence, especially in PoS systems [17].

The regulatory environment also shapes consensus power distribution. In jurisdictions with complex or restrictive regulations, only well-funded entities can afford to comply, limiting participation to larger players capable of navigating regulatory challenges effectively [18].

Incentive structures within blockchain networks can unintentionally contribute to power concentration. In PoW systems, reward mechanisms may disproportionately benefit large mining pools, while delegation models in PoS networks often concentrate power among larger validators [19].

Addressing these factors is essential for designing blockchain networks that minimise consensus power inequality. Strategies such as reducing economic barriers, ensuring fair token distribution, adopting decentralised governance models, and revising incentive structures can promote more equitable participation. These measures enhance network security, uphold decentralisation, and foster the healthy development and adoption of blockchain ecosystems. By tackling the underlying causes of consensus power inequality, blockchain developers and stakeholders can work towards achieving a more balanced distribution of power. This approach ensures network integrity and aligns with the foundational principles of blockchain technology, safeguarding its potential as a transformative, decentralised solution.

4. Importance of Distributed Consensus Power

Decentralisation is a foundational principle of blockchain technology, aimed at eliminating central points of control and failure. A key aspect of achieving this decentralisation is the distribution of consensus power, which ensures that decision-making authority is not concentrated in the hands of a single entity or a small group. By evenly distributing influence over the consensus mechanism, blockchain networks maintain their decentralised nature and uphold the trust of their participants [4].

The distribution of consensus power plays a critical role in enhancing the security and robustness of blockchain networks. When consensus power is widely dispersed among participants, the likelihood of malicious activities, such as 51% attacks, is significantly reduced. In such an attack, a single entity or coalition would need to control a majority of the network's computational or staking power to alter transactions or manipulate the blockchain. A more distributed power structure mitigates this risk, thereby preserving the integrity and reliability of the network [20]. This aspect is particularly vital for public blockchains, where open participation makes them more susceptible to exploitation if consensus power becomes concentrated.

Beyond security, distributed consensus power enhances the trustworthiness and credibility of blockchain networks. Users and developers are more inclined to engage with networks that operate transparently and equitably, knowing that decision-making processes are not dominated by a select few. A fair distribution of power fosters a sense of inclusivity and ownership among participants, creating an environment where contributions from a diverse range of stakeholders drive network growth and sustainability.

Furthermore, distributed consensus power encourages broader participation and innovation within blockchain ecosystems. By lowering entry barriers, it allows individuals and smaller entities to actively participate in consensus processes without being overshadowed by large, well-funded participants. This inclusivity not only democratises blockchain participation but also promotes diversity in ideas and solutions, driving the continuous improvement and evolution of blockchain systems.

## 5. Data and Methodology

To evaluate the distribution of consensus power across various blockchain networks, we collected data on computational power, staking, and other consensus-related resources allocated among network participants. The dataset spans three years, from January 2022 to July 2024, and was analyzed using metrics such as the Gini and Theil coefficients to assess the level of decentralisation. For Hedera, data collection began on 27 July 2022, marking the introduction of dynamic staking and the point at which it became meaningful to analyze the distribution of consensus power among its nodes.

Mining and staking entities for Bitcoin and Ethereum were identified using public data explorers such as Etherscan [21], Dune [22], Stakingrewards.com [23] and Blockchain.com [24]. For Bitcoin, hashrate distribution data across 27 labeled mining pools and miners was collected, covering an average market share of 81.2% during the analysis period from January 2022 to July 2024. For Ethereum, data collection spanned both its PoW and PoS periods. During the PoW phase (January 2022 to September 2022), hashrate distribution across 44 labeled mining pools and miners was captured, representing an average market share of 89.3%. During the PoS phase (September 2022 to July 2024), data on staked tokens across 100 labeled staking pools and stakers was gathered, with an average market share of 84.5%. Table 1 summarises the data sources used to evaluate the distribution of consensus power across the five blockchain networks analysed in this study.

For PoS blockchains, we specifically analysed the distribution of staked amounts across pools and stakers. For the Bitcoin PoW and Ethereum PoW, we focused on the distribution of hashrate power across mining pools and miners. However, it is important to note that applying this analytical framework to certain blockchains presents challenges. For example, the XRP Ledger (XRPL) uses a unique consensus algorithm where power is distributed among trusted validators in one or more Unique Node Lists (UNLs). While participants can propose and use different UNLs, the actual configuration is not publicly disclosed by server operators, making it difficult to measure consensus power distribution. Although the XRPL Foundation updates its recommended UNL, this does not provide a full picture of UNL usage across the network [25]. Due to these data limitations, XRPL was excluded from our analysis

| DLT | Consensus | Type of Data | Data | Period |
| --- | --- | --- | --- | --- |
| Bitcoin | PoW | Hashrate per miner | 27 mining pools and miners | Jan 2022 – July 2024 |
| Ethereum | PoW | Hashrate per miner | 44 mining pools and miners | Jan 2022 – Sep 2022 |
| Ethereum | PoS | Staked tokens per staker | 100 staking pools and stakers | Sep 2022 – July 2024 |
| Cardano | PoS | Staked tokens per staker | 2,648 staking pools and stakers | Jan 2022 – July 2024 |
| Hedera | PoS | Staked tokens per consensus node | 32 consensus nodes | July 2022 – July 2024 |
| Algorand | PoS | Staked tokens per staker | 18,200 staker addresses | Jan 2022 – July 2024 |

Table 1. Consensus power data sources.

To estimate the inequalities across the network we have relied on two main indicators: Gini coefficient [26] and Theil index [27].

The Gini coefficient is a widely used metric in economics to measure income inequality within a population. It evaluates inequality by analyzing the distribution of income levels across the population. Mathematically, the Gini coefficient is defined as half the mean absolute difference between all pairs of elements in the population, normalised by the population's average income. Here, $N$ represents the population sise, and $x_i$ denotes the income (or property) of element $i$.

$$G = \frac{\sum_{i=1}^{N}(2i - N - 1) \cdot x_i}{N \cdot \sum_{i=1}^{N} x_i} \quad (1)$$

In our analysis, we use the inverted Gini coefficient, where a value of 0 indicates perfect inequality and 1 represents perfect equality. This inversion allows for a more intuitive interpretation in the context of our study.

The Theil index, on the other hand, is another measure of economic inequality that is also applicable in information theory to represent redundancy. It quantifies inequality by comparing the maximum possible diversity of data with the observed diversity. This metric highlights the degree of inequality and distribution within a dataset.

$$T = \frac{1}{N} \sum_{i=1}^{N} \left(\frac{x_i}{\mu}\right) \ln\left(\frac{x_i}{\mu}\right) \quad (2)$$

To align the Theil index with the inverted Gini coefficient for comparison purposes, we have also normalised the Theil index to a range of 0 to 1, where 0 indicates a lack of equality, and 1 represents perfect equality.

We applied the aforementioned metrics to analyse the distribution of consensus power among participants—such as miners or validators—within a blockchain network. This assessment evaluates the degree of disparity in consensus power among network participants. In our calculations, $N$ represents the total number of consensus participants, $x_i$ denotes the consensus power value (e.g., staked tokens or hashrate) of the $i$-th participant, and $\mu$ is the mean of the $x_i$ values.

These metrics are critical for identifying imbalances where a small subset of participants controls a disproportionately large share of the consensus power, signifying an unequal distribution as the metric value approaches 0. Conversely, when consensus power is evenly distributed across participants, indicating a lack of concentration, the metric values are close to 1.

## 4. Results

The Gini coefficient analysis (Figure 1) reveals varying levels of consensus power equality among the five blockchain networks. Hedera exhibits slight fluctuations but consistently remains at the higher end of the equality spectrum, with its inverted Gini coefficient close to 0.9 throughout the observed period, indicating an evenly distributed consensus power among its nodes. Bitcoin demonstrates intermediate levels of equality, with its inverted Gini coefficient fluctuating between 0.4 and 0.8. A significant decline is observed after mid-2022, stabilising just below 0.4 from the beginning of 2023. Ethereum's inverted Gini coefficient experienced a significant decline following the merge on 15 September 2022, dropping from 0.3 to 0.15. Cardano exhibits a high concentration of consensus power, maintaining a stable value just below 0.15 throughout the observed period. In contrast, Algorand demonstrates the lowest inverted Gini coefficient, remaining close to zero during the first half of the study, with a slight upward trend emerging from mid-2023. These findings underscore the stark differences in consensus power dynamics across the networks, as highlighted by the Gini coefficient.

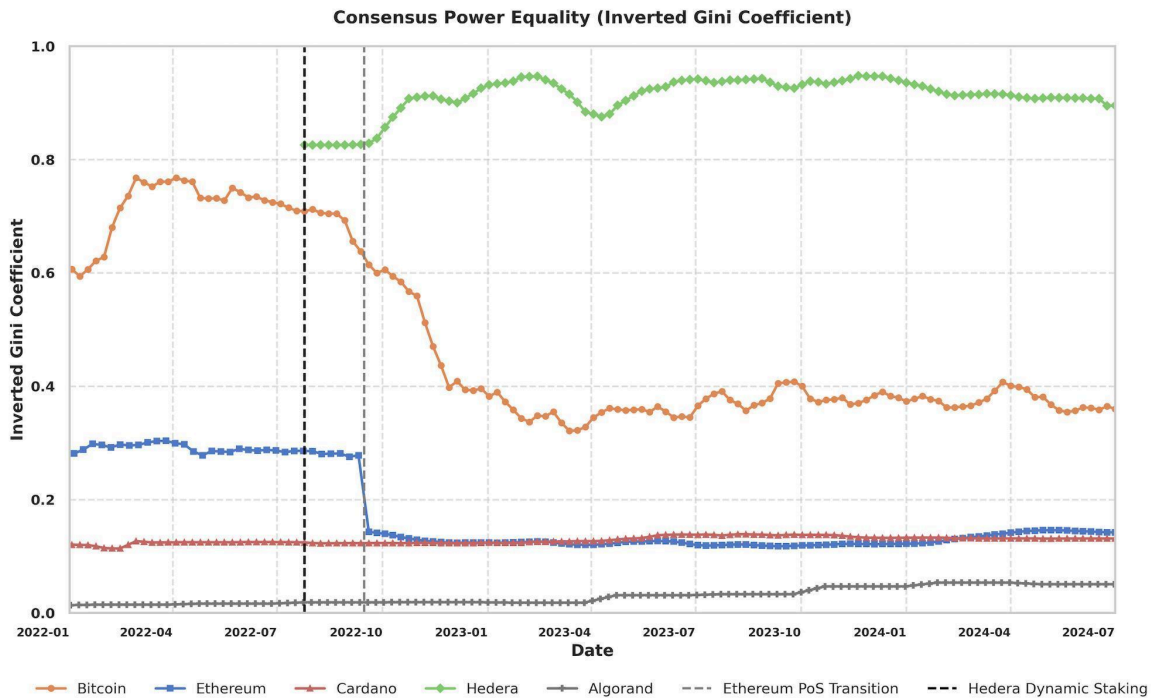

Figure 1. Consensus power equality across blockchain networks (Gini coefficient, Jan 2022 – July 2024). A Gini coefficient closer to 1 indicates greater equality, while a value closer to 0 reflects more concentrated power.

The Theil index analysis (Figure 2) complements the Gini coefficient by providing a more nuanced perspective on relative equality in consensus power distribution. Hedera stands out with a consistently high inverted Theil index value, remaining near 1 throughout the observed period. Bitcoin begins with a high inverted Theil index in 2022, which gradually declines to moderate levels, mirroring trends observed in the Gini coefficient. Cardano exhibits stable but comparatively lower equality levels than Bitcoin, with minimal variation over time. Ethereum shows moderate consensus power equality during its PoW phase, with an inverted Theil index stable at 0.6. However, this drops to 0.4 following the merge, reflecting increased concentration. Algorand, in contrast, records the lowest inverted Theil index values but displays a slight upward trend starting in mid-2022. By quantifying both redundancy and diversity in power distribution, the Theil index provides complementary insights to the Gini coefficient, highlighting subtle differences across the blockchain networks.

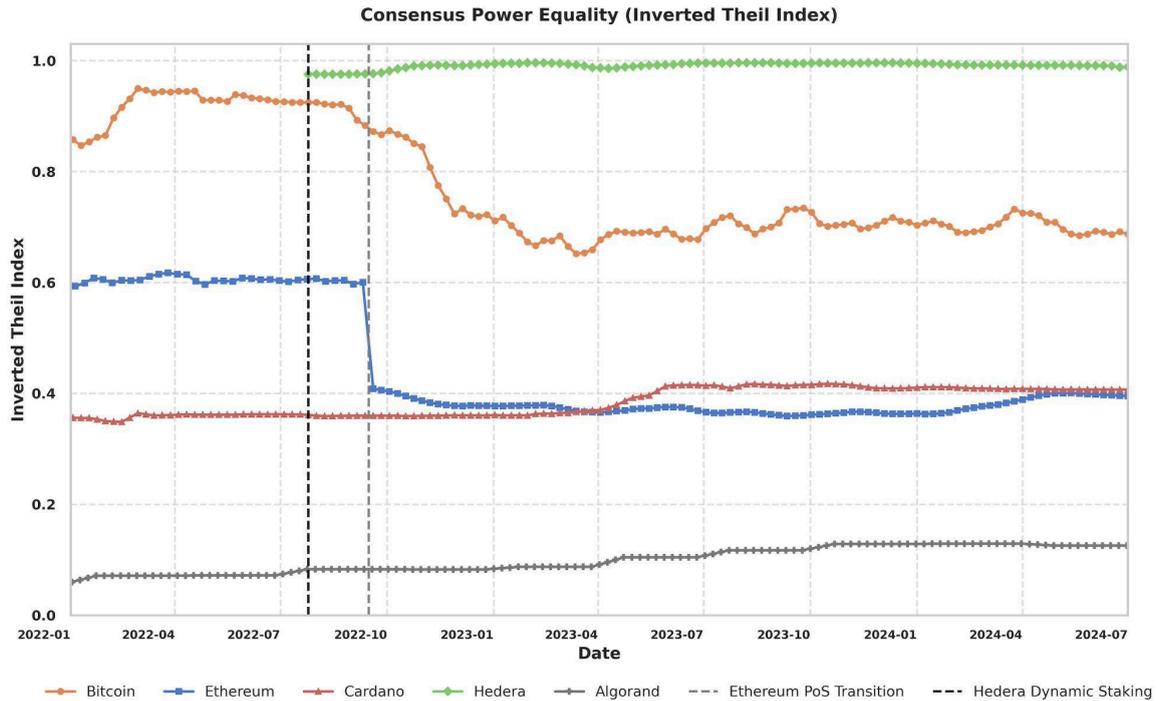

Figure 2. Consensus power equality across blockchain networks (Theil index, January 2022 – July 2024). A higher Theil index indicates a more equal distribution of consensus power, whereas a lower index reflects greater concentration.

## 6. Discussion

From January 2022 to July 2024, the trends in consensus power equality across the sampled blockchains have remained relatively stable. Among the networks analysed, Hedera consistently demonstrates the most equal distribution of consensus power. This stability is attributed to Hedera's unique governance structure and the implementation of dynamic staking on July 27 2022, which actively promotes the equitable distribution of consensus power among nodes. In contrast, Bitcoin shows the highest volatility in consensus power equality. The decline observed in 2022 reflects the dominance of a few large mining pools inherent to Bitcoin's proof-of-work model, which contributes significantly to inequality in power distribution. Ethereum and Cardano exhibit intermediate levels of equality, with Ethereum experiencing a decline in consensus power equality following its transition to proof-of-stake on September 15, 2022. While Algorand has a broad base of consensus participants, its power distribution remains concentrated among a select few. This is evident in Algorand's consistently lower Gini and Theil scores compared to the other blockchains. These results suggest that a small group of participants, holding significant amounts of the native cryptocurrency, exerts disproportionate influence over the network, overshadowing smaller stakers and limiting decentralisation efforts.

Table 2 provides a detailed comparison of the rankings of the blockchains based on consensus power equality. Hedera ranks first, owing to its Hashgraph consensus mechanism, which facilitates a more equitable distribution of power among participants. Bitcoin follows in second place, maintaining a relatively balanced distribution despite the persistent dominance of a few major mining pools within its proof-of-work framework. Ethereum ranks third, showing a decline in equality following its transition to proof-of-stake. Cardano, utilising the Ouroboros proof-of-stake protocol, exhibits a comparable distribution pattern to Ethereum, with intermediate levels of equality. Finally, Algorand ranks last, as its consensus power remains heavily concentrated among a small number of participants, a characteristic that is reflected in its low Gini and Theil equality scores.

The inclusion of both public and public-permissioned blockchains in the analysis offers a well-rounded perspective on consensus power distribution across various blockchain architectures. Both public and public-permissioned blockchains allocate consensus power to network participants—whether through open competition, as in public blockchains, or through semi-restricted mechanisms, as in public-permissioned ones.

| Rank | Blockchain | Consensus Mechanism | Consensus Power Distribution |
|---|---|---|---|
| 1 | Hedera | Hashgraph Consensus | Hedera's consensus model adopts a distinctive approach designed to promote an equitable distribution of power across the network. |
| 2 | Bitcoin | Proof-of-Work | Bitcoin mining is less concentrated than it might appear, despite the top mining pools controlling a substantial portion of the hash rate. |
| 3 | Ethereum | Proof-of-Work / Proof-of-Stake | During its PoW phase, Ethereum exhibited moderate consensus power equality; however, following the Merge, consensus power distribution became more concentrated, reflecting increased reliance on large staking pools. |
| 4 | Cardano | Proof-of-Stake | Cardano's Ouroboros protocol demonstrates an uneven distribution of consensus power, comparable to that of Ethereum's PoS network. |
| 5 | Algorand | Pure Proof-of-Stake | Algorand's consensus power distribution is highly concentrated, with a small group of participants exerting considerable influence over the network's consensus process. |

Table 2. Ranking of consensus power equality, including details on the consensus mechanism and its distribution among network participants.

It is important to emphasise that this analysis focuses exclusively on the distribution of consensus power and does not suggest that any network is more decentralised than Bitcoin overall. Decentralisation is a multifaceted concept encompassing various dimensions beyond consensus power, such as governance structures, network accessibility, and protocol control mechanisms. While Bitcoin operates within a permissionless framework, it exhibits higher concentration in terms of consensus power compared to others, as shown by the metrics used in this study. This focused analysis provides a specific lens on consensus power distribution while acknowledging the broader complexity of decentralisation.

7. Conclusion

The distribution of consensus power is a critical factor in evaluating the health and sustainability of blockchain networks. Our comparative analysis reveals varying degrees of inequality among the top blockchain networks. Hedera and Bitcoin demonstrate more balanced power distribution, aligning closely with the principles of decentralisation. In contrast, Ethereum, Cardano, and Algorand show higher levels of concentration, primarily due to the influence of large validators or stakers who hold a significant portion of the network's resources. This concentration can limit decentralisation, as a smaller group of participants retains a disproportionate amount of control over the consensus process, diminishing the influence of smaller participants and reducing the overall balance in power distribution.

As blockchain technology continues to evolve, maintaining a focus on distributed consensus power will be essential for fostering secure, decentralised, and innovative ecosystems across all networks. Future research and monitoring will be crucial to ensure that these networks continue to comply with the foundational principles of blockchain.